\documentclass[runningheads]{llncs}
\usepackage{graphicx}
\usepackage{amsmath}
\usepackage{amssymb}
\usepackage{wrapfig}
\usepackage{caption}
\begin{document}
\title{MixCL: Pixel label matters to contrastive learning in medical image segmentation}
%
\titlerunning{MixCL: Pixel label matters to contrastive learning}
%
\author{Jun Li\inst{1} \and Quan Quan\inst{1} \and 
S. Kevin Zhou\inst{1,2}}
%
\institute{Key Lab of Intelligent Information Processing of Chinese Academy of Sciences (CAS),
Institute of Computing Technology, CAS, Beijing, 100190, China \and School of Biomedical Engineering \& Suzhou Institute for Advanced Research Center for Medical Imaging, Robotics, and Analytic Computing \& LEarning (MIRACLE), University of Science and Technology of China, Suzhou 215123, China 
}

\maketitle   
\begin{abstract}
Contrastive learning and self-supervised techniques have gained prevalence in computer vision for the past few years. It is essential for medical image analysis, which is often notorious for its lack of annotations. Most existing self-supervised methods applied in natural imaging tasks focus on designing proxy tasks for unlabeled data. For example, contrastive learning is often based on the fact that an image and its transformed version share the same identity. However, pixel annotations contain much valuable information for medical image segmentation, which is largely ignored in contrastive learning. In this work, we propose a novel pre-training framework called Mixed Contrastive Learning (MixCL) that leverages both image identities and pixel labels for better modeling by maintaining identity consistency, label consistency, and reconstruction consistency together. Consequently, thus pre-trained model has more robust representations that characterize medical images. Extensive experiments demonstrate the effectiveness of the proposed method, improving the baseline by 5.28\% and 14.12\% in Dice coefficient when 5\% labeled data of Spleen and 15\% of BTVC are used in fine-tuning, respectively.

\keywords{Constrastive learning \and Self-supervision \and Pre-training.}
\end{abstract}
\section{Introduction}
Self-Supervised Learning (SSL) obtains supervisory signals from the data itself without any human annotations.
Substantial progress has been made in image classification~\cite{pathak2016context,gidaris2018unsupervised,he2020momentum,zbontar2021barlow}, object detection~\cite{pathak2016context,yao2021one}, and semantic segmentation~\cite{gidaris2018unsupervised,shimoda2019self,zhu2020rubik,zheng2021hierarchical,zhang2021self,liu2020self}. Many works also prove its superiority on  various medical tasks~\cite{windsor2021self,hu2021self,yang2021self,xu2021deformed2self,sahasrabudhe2020self,lu2020white,bhalodia2020self}.
Data is the footstone of deep learning. Likewise, the remarkable promotion of SSL comes from exploiting a large amount of unlabeled data. 

One practical way of curating such `big data' is through aggregating multiple datasets. From the perspective of data utilization, SSL leverages unlabeled data in a task-agnostic fashion as the supervised labels are only used during fine-tuning. Even though these datasets may contain pixel-wise annotations, these annotations are largely ignored in SSL pre-training. In this paper, we aim to bring them back for improved SSL pre-training in the context of medical image segmentation, reaching full exploitation of images and annotations at hand.


SSL involves designing a framework that pre-trains a model from $n$ fully annotated datasets $\mathcal{D}=\{D_0,D_1,\dots,D_n\}$, which, after adaptation, likely achieves better performance on target task $D_{t}$. Mathematically, it can be depicted as:
\begin{equation}
\mathop{\arg\max}_{\theta} \ \mathcal{M} (f(x|\theta),y) \wedge \mathop{\arg\min}_{\theta} \ \mathcal{L} (f(x^{\prime}, y^{\prime}|\theta )), \ \  [x, y] \in D_t, \  [x^{\prime}, y^{\prime}] \in \mathcal{D} \label{problem defination},
\end{equation}
where $\mathcal{M}(\cdot)$ is the performance metric on target task $D_{t}$, and $\mathcal{L}(\cdot)$ is the objective function for pre-training.  

The annotations from different datasets provide diverse observations of images from varying aspects, potentially helping build models for improved image understanding.~\cite{zhou2021review} However, we often leverage these annotations in a task-specific way~\cite{zhou2019handbook}. It leaves the training process in a muddle when multiple annotations from $\mathcal{D}$ get mixed up, and this is where the difficulty of Problem~\eqref{problem defination} lies. Namely, an appropriate design of $\mathcal{L}(\cdot)$ should keep gradient directions consistent when training by a hotchpotch of all kinds of datasets.

Contrastive Learning (CL) has achieved a remarkable level of performance~\cite{chen2020simple,chen2020big,he2020momentum,grill2020bootstrap,zbontar2021barlow}. It aims to learn invariant representations under different data augmentations. The representations of paired similar samples stay close while dissimilar ones are far apart. The relation of ``similar-dissimilar" is based on image-level comparison as in SimCLR~\cite{chen2020simple}, BYOL~\cite{grill2020bootstrap} and Barlow Twins~\cite{zbontar2021barlow}. It is extended to pixel-level for dense downstream tasks like segmentation~\cite{wang2021exploring,xie2021propagate}. When viewing medical images in $\mathcal{D}$ at pixel-level, the similar and dissimilar pairs could be defined by the annotations, naturally introduced as the supervised signal. This kind of supervised signal is not hard and strict but soft and mild. On the other hand, images in $\mathcal{D}$ are with labels of saying different organs or lesions, which can be viewed as tissues and fluid in the human body from a fine-grained perspective. Pixel-level CL has the potential to learn the compatibility among diverse labels.

Inspired from the above discussions, we propose a novel contrastive learning framework named Mixed Contrastive Learning (MixCL) to give a solution to Problem~\eqref{problem defination}. The proxy task for MixCL has three components: the \textbf{reconstruction consistency} and \textbf{identity consistency} terms are designed to fully utilize image; the \textbf{label consistency} term is introduced to exploit labels as supervisory signal and ease the negative influence by the diversity of annotations in pre-training phase. The proposed MixCL achieves remarkable improvement on the downstream segmentation task defined in the BTCV and MSD Spleen datasets.

\begin{figure}
\centering
\includegraphics[width=1.0\textwidth]{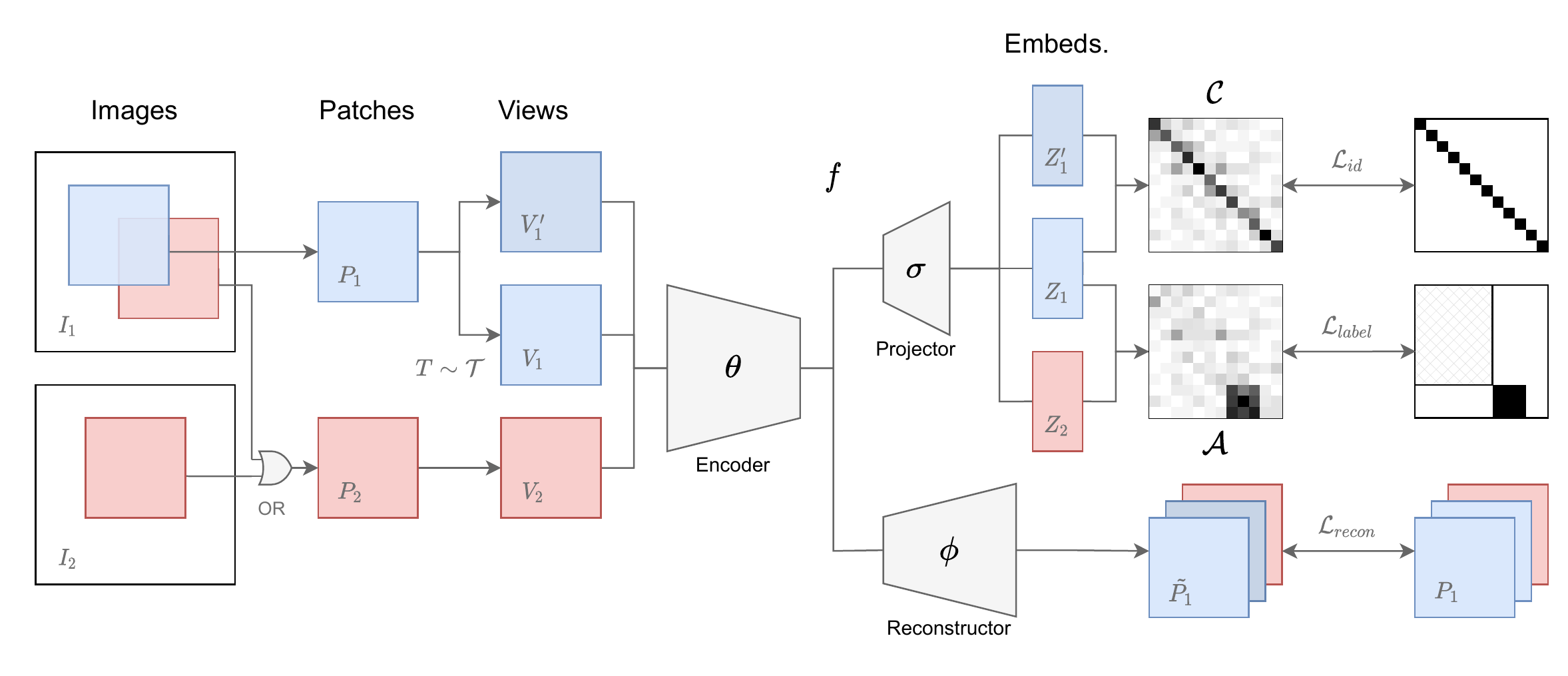}
\caption{Overview of the proposed Mixed Contrastive Learning (MixCL). } \label{fig:framework}
\end{figure}

\section{Method}

\subsection{Overview of MixCL}
\label{sec:Overview}

We employ UNETR~\cite{hatamizadeh2022unetr} as the backbone network, which is one of the most successful Vision Transformers (ViT)s~\cite{dosovitskiy2020image} in medical image analysis. As shown in Fig.~\ref{fig:framework}, the encoder $\theta$ and reconstructor $\phi$ form a typical UNETR, having the same settings in~\cite{hatamizadeh2022unetr}. The projector $\sigma^l$ corresponding to the $l$-th layer of encoder is composed of three $1\times1\times1$ convolution layers with channels set as [$c^l, c^l, 2\times c^l$], interspersed with normalization and activation layers.

A patch $P_1$ is first cropped from an original image $I_1 \in \mathcal{D}$ with corresponding segmentation $S_1$, and then its two different views $V_1$ and $V^{\prime}_1$ are obtained by random augmentations $\mathcal{T}$. $V_1$ and $V^{\prime}_1$ are then fed into the UNETR encoder with parameters $\theta$, followed by a projection head $\sigma$ to produce embeddings $Z_1$ and $Z^{\prime}_1$, respectively. A reconstruction decoder $\phi$ also follows the encoder $\theta$ to generate restored patches $\tilde{P_1}$ and $\tilde{P^{\prime}_1}$. Simultaneously, an auxiliary patch $P_2$ with its label $S_2$ is treated the same as $P_1$, acquiring $Z_2$ and $\tilde{P_2}$. Details about $P_2$ would be introduced in Section~\ref{sec:LabelConsistency}. 

Note that $Z_1$ is a projection set over all intermediate outputs of each layer in the encoder $\theta$, and $Z^1_1$ denotes the projection vector of the 2-nd layer. The \textit{identity consistency loss} $\mathcal{L}_{id}(\cdot)$ is applied on all intermediate layers by calculating the cross-correlation matrix between $Z_1$ and $Z^{\prime}_1$. $\mathcal{L}_{id}(\cdot)$ together with the \textit{reconstruction loss} over $\tilde{P_i}$ and ${P_i}$,  mines knowledge via image itself.
Besides, \textit{label consistency loss} is applied on the 2-nd layer over $Z^1_1$ and $Z^1_2$, which is a supervised contrastive loss utilizing annotations appropriately. 
The proposed framework of MixCL is formulated as follows: 
\begin{equation}
    \mathcal{L} = \mathcal{L}_{id}(Z_1, Z^{\prime}_1|\theta, \sigma) + \alpha \mathcal{L}_{label}(Z^1_1, Z^1_2|\theta, \sigma) + \beta \mathcal{L}_{recon}(P_i, \tilde{P_i}|\theta, \phi),
\end{equation}
where $\alpha, \beta$ are scale factors, and both are set to 0.01 by default. The identity consistency loss $\mathcal{L}_{id}(\cdot)$ and label consistency loss $\mathcal{L}_{label}(\cdot)$ will be explaine in Section~\ref{sec:IDConsistency} and Section~\ref{sec:LabelConsistency}, respectively, in detail.
The reconstruction consistency between $\tilde{P_i}$ and $P_i$ is given as
\begin{equation}
\mathcal{L}_{recon}(P_i, \tilde{P_i}|\theta, \phi) = \|P_i-\tilde{P_i}\|^2.
\end{equation}

\subsection{Identity Consistency} \label{sec:IDConsistency}

Identity consistency is employed to maximize the similarity between the representations  $Z_1$ and $Z^{\prime}_1$, obtained from two distorted versions of sample $P_1$. In vanilla Barlow Twins~\cite{zbontar2021barlow}, $Z_1$ and $Z^{\prime}_1$ embed image-level information into vectors. Here, we transfer it from instance-level to pixel-level. The shape of $Z_1$ is $[L, B\times H\times W \times D, C]$, which means that all pixels are embeded into a $C$-dimensional space. Then, the $\mathcal{L}_{id}(\cdot)$ is defined as :
\begin{equation}
    \mathcal{L}_{id}(\mathcal{C}|\theta, \sigma) = \sum_l \left [\sum_i{{(1-\mathcal{C}_{ii}^l)}^2} + \lambda \sum_i\sum_{j\neq i}{\mathcal{C}_{ii}^l}^2 \right ],
\end{equation}
where $\lambda$ is a balance factor, and $\mathcal{C}^l$ is the cross-correlation matrix computed by $Z_1^l$ and ${Z^l_1}^{\prime}$ along dimension 0 at the $l$-th layer:
\begin{equation}
    \mathcal{C}_{ij}^l \triangleq \frac {\sum_M Z^l_{M,i} {Z^l_{M,j}}^{\prime} }
    {\sqrt{\sum_M {(Z^l_{M,i})}^2}\sqrt{\sum_M {({Z^l_{M,j}}^{\prime})}^2}}, \ \ M = B\times H \times W\times D,
\end{equation}
where $M$ is the total number of pixels, and $i$ and $j$ are the index for a vector in the  $C$-dimensional space.

\begin{wrapfigure}[23]{r}{0.5\textwidth}
  \centering
  \includegraphics[width=0.5\textwidth]{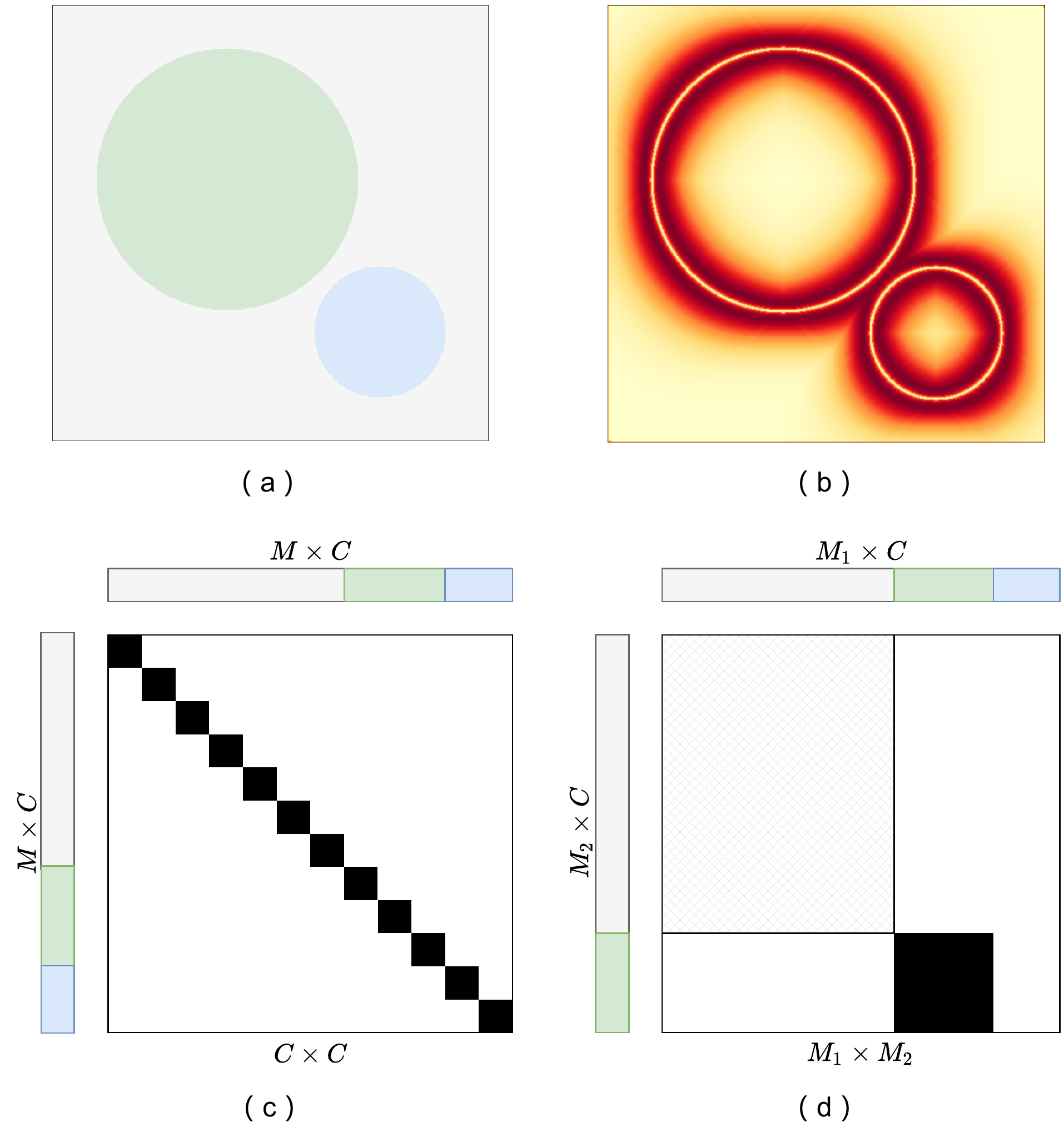}
\caption{Schematic diagrams of (a) Segmentation map with 2 organs labeled in blue and green, background in gray; (b) Weight map for pixel sampling; (c) The cross-correlation matrix $\mathcal{C}$ in identity consistency; and (d) The label relation matrix $\mathcal{A}$.} 
  \label{fig:WeightMapLossMatrix}
\end{wrapfigure}

As shown in Fig.~\ref{fig:WeightMapLossMatrix} (c), the diagonal elements of the cross-correlation matrix $\mathcal{C}$ to be 1 in the first term of $\mathcal{L}(\cdot)$, making each unit of embeded features invariant over all pixels. Whereas the second term aims to de-correlate representation components, the off-diagonal elements are set to be 0. It can be viewed as a soft-whitening constraint on embedding space, reducing redundancy. The $\mathcal{L}_{id}(\cdot)$ is capable of encoding rich image information into model parameters, simple and easy to implement. 

\subsection{Label Consistency}\label{sec:LabelConsistency}
Indeed, all information derives from images, including the annotations created manually.
Things get complicated when training all kinds of annotations at once.
The gradient directions to various labels are not consistent, failing the training process. Consequently, the supervisory signals from different labels should be consistent in solution space, denoted as Label Consistency. This section will introduce the label consistency loss to tackle this issue.

\subsubsection{Auxiliary Sample}
MixCL attempts to utilize the mixed annotations in a way that satisfies the Label Consistency. A patch $P_2$ is picked to assist MixCL achieving this purpose. As shown in Fig.~\ref{fig:framework}, $P_2$ is randomly cropped from image $I_1$ or $I_2$ at a 1:1 ratio. Notice that $I_1$ and $I_2$ come from the same dataset $D_i \in \mathcal{D}$. As mentioned in Section~\ref{sec:Overview}, $P_2$ is embeded into $Z_2$, which is going to compute $\mathcal{L}_{label}(\cdot)$ with $Z_1$ in 2-nd layer, taking labels into account.

\subsubsection{Objective Function}
The fundamental requirement of $\mathcal{L}_{label}(\cdot)$ is guiding the training process with the favor of annotations. Fig.~\ref{fig:WeightMapLossMatrix} (a) shows a schematic illustration of segmentation from $D_i \in \mathcal{D}$. Saying ‘liver’ in green, ‘liver tumor’ in blue and 'background' in grey. With the two feature maps $Z^1_1$ and $Z^1_2$ computed from two different patches $P_1$ and $P_2$, pixels labeled by 'liver' in $Z^1_1$ and those with the same label in $Z^1_2$ should be regarded as 'Positive-pairs'. And different labeled pixels in $Z^1_1$ and $Z^1_2$ is treated as 'Negative-pairs'. This is the label consistency at images-level.

Fig.~\ref{fig:WeightMapLossMatrix} (d) shows the label relation matrix $\mathcal{A}$, where positive pixels pairs are set to be 1 (black area), and negative pixels pairs are set to be 0 (white area). The shadow area denotes background pixel pairs over pixels indicating 'kidney', 'pancreas' or other labels in rest datasets. The background pixel pairs are not considered in Mixel-CL, for the sake of complex composition of background pixels would break the compatibility between different datasets. This is the label consistency at datasets-level.

Then, the contrastive task could be formulate as:
\begin{equation}
    \mathcal{L}_{label}(Z^1_1, Z^1_2|\theta, \sigma) = -\log \frac{\sum\limits_{(i,j)\in [\mathcal{A}=1]} \exp{({\cos(Z^1_{1,i}, Z^1_{1,j})/\tau})}}{\sum\limits_{(i,k)\in [\mathcal{A}=0]\cup [\mathcal{A}=1]} \exp{({\cos(Z^1_{1,i}, Z^1_{1,k})/\tau})}},
    \label{equ:mixed}
\end{equation}
where $\tau$ is a temperature hyper-parameter. All of our results use a temperature of $\tau=0.1$. $[\mathcal{A}=0]$ and $[\mathcal{A}=1]$ represent the pixel pairs assigned as negative and positive in label relation matrix $\mathcal{A}$, respectively.

\subsubsection{Pixel Sampling}
However, the pixel-wise computation of $\mathcal{L}_{label}(\cdot)$ requires $O(n^2)$ complexity of time and space consuming. We adopt a trade-off strategy, sampling an appropriate amount of pixels from $Z^1_1$ and $Z^1_2$. Given a budget of $r$ pixels for each segmentation class, the sampling is performed on a weight map $W(\cdot)$, defined as:
\begin{equation}
    D(S) = \inf\{d(i,j):j\in \partial S\}, i \in S^{\circ},~
    W(S) = \frac{1}{2}\sqrt{D(S)}\exp(\frac{-D(S)}{\mu}) + \epsilon,
\end{equation}
where $\partial S$ is the boundary of each label and $S^{\circ}$ is the interior. Then $D(S)$ computes the distance of the interior pixels to its boundary. $\mu$ and $\epsilon$ are two hyper-parameters, scaling the probabilities of the weight map. The default $\mu$ and $\epsilon$ are set to 8 and 0.05. As shown in Fig.~\ref{fig:WeightMapLossMatrix} (b), interior pixels closer to the boundary are more discriminative, with higher values assigned, therefore.

Going this far on $\mathcal{L}_{label}(\cdot)$, pixel-wise annotations of all datasets have been exploited. It keeps the same numerical form among different datasets. The complex background, the main destabilizing factor, is also duly handled. Both image-level and datasets-level label consistency ensure a clear gradient descent direction. By applying MixCL, downstream tasks gain sizable performance boosts.

\section{Experiments}



\subsection{Implementation Details}
\label{sec:datasets}

\textbf{Datasets} Up to 765 CT scans with annotations are used for pre-training, which come from (1) Medical Segmentation Decathlon (MSD) dataset~\cite{antonelli2021medical} (only Liver, Lung and Pancreas are used for pre-training), (2) NIH Pancreas-CT~\cite{roth2016data}, and (3) 2019 Kidney Tumor Segmentation Challenge (KiTS)~\cite{heller2019state}. The pre-trained model is then fine-tuned on two datasets: The Beyond the Cranial Vault (BTCV)~\footnote{https://www.synapse.org/\#!Synapse:syn3193805/wiki/89480} and Spleen segmentation in MSD. BTCV provides 30 CT scans with 13 abdominal organs annotated, while Spleen segmentation in MSD has 41 annotated CT volumes.


\label{sec:Implementation}
\noindent\textbf{Augmentation}
Each CT scan in datasets are interpolated to a voxel spacing of $1.5 mm\times1.5 mm\times 2.0 mm$, with intensities scaled to $[-150, 250]$. 
For the pre-training task, random elastic deform and affine transform are applied on $96\times96\times96$ 3D patches, both under the probability of 0.5. Then, patches are distorted into views via random intensity range shifting, random Gamma transformation, random Gaussian smoothing and noising, which have the same probability of 0.2.
Augmentations like random rotation in 90, 180 and 270 degrees, random flip in axial, sagittal and coronal views and random scale and shift are used in the fine-tuning tasks, with the same probability of 0.2.

\noindent\textbf{Optimization}
We use the AdamW optimizer~\cite{loshchilov2017decoupled} with a cosine learning rate scheduler. The learning rate is set to $1e-4$ initially using a decay of $1e-5$ for 1000 iterations in pre-training and 2000 iterations in fine-tuning. Pre-training experiments use a batch size of 1 per GPU across 6 RTX Titan, and fine-tuning uses a batch size of 2. 

\begin{table}[t]
    \centering
    \caption{Performance comparison in terms of Dice coefficient between the target models initialized with or without our pre-trained model. `Pct.' refers to the percentage of labeled data used in fine-tuning. `T.F.S.' stands for `train from scratch'.}
    \setlength{\tabcolsep}{1.6mm}
    \begin{tabular}{rrrr||rrrr}
        \hline
        \multicolumn{4}{c||}{Spleen Dataset} & \multicolumn{4}{c}{BTVC Dataset}\\
        \hline
        Pct. & T.F.S & Ours & Diff.&  Pct. & T.F.S. & Ours & Diff. \\
        \hline
        5\% & 0.5065 & 0.5593 & +0.0528& \\
        15\% & 0.8247 & 0.8430 & +0.0183 & 15\% & 0.2459 & 0.3871 & +0.1412\\
        50\% & 0.9424 & 0.9453 & +0.0019 & 50\% & 0.5851 & 0.5939 & +0.0085  \\
        100\%  & 0.9597 & 0.9612 & +0.0015 & 100\% & 0.7524 & 0.7656  & +0.0132\\
        \hline
    \end{tabular}
    \label{table:main}
\end{table}

\subsection{Results}

\textbf{Quantitative performances.}
In the fine-tuning process, the encoder of UNETR~\cite{hatamizadeh2022unetr} is initialized by the pre-trained weights. The effectiveness of MixCL is assessed by the performance on a 5-fold cross-validation on fine-tuning tasks. As mentioned in Section.~\ref{sec:datasets}, we fine-tune the pre-trained model on MSD Spleen and BTCV, and compare the Dice metric with models trained from scratch under the same training settings. In addition, comparisons on partially labeled data are also performed, giving as assessment from varying perspectives. The experiments and results are shown in Table~\ref{table:main}. 

Our method achieves a consistent performance lift on both datasets, and such a lift is more pronounced when less labeled data is used in fine-tuning. For example, the Dice coefficient is improved by 5.28\% in the Spleen dataset (5\% data) and by 14.12\% in the BTVC dataset (15\% data), yet such an improvement in Dice is 0.15\% in the Spleen dataset (100\% data) and 1.32\% in the BTVC dataset (100\% data). Another interesting observation is that the use of pre-trained model seems to bring more improvements to a complicated task, that is, segmentation of 13 organs, than to a simple task of spleen segmentation. These improvements demonstrate that MixCL successfully exploits mismated labels all at once in contrastive learning, and obtains informative and discriminative representations.

\begin{table}[t]
    \centering
    \caption{Ablation study on three components of MixCL performed on MSD Spleen with 30\% labeled data, and BTCV with 100\% labeled data . Id: identity consistency, Label: label consistency, Recon: image reconstruction. }
    \setlength{\tabcolsep}{1.6mm}
    \begin{tabular}{lrcccr}
        \hline
        Dataset & Pct. & Identity & Label & Recon & Dice \\
        \hline
        Spleen & 30\% &           &            &            &   0.8975 \\
               &     &\checkmark & \checkmark &            &   0.9150 \\
               &     &\checkmark & \checkmark & \checkmark &   0.9178 \\
        \hline
        BTCV   & 100\% &           &            &            &   0.7524 \\
               &     &\checkmark & \checkmark &            &   0.7618 \\
               &     &\checkmark & \checkmark & \checkmark &   0.7656 \\
         \hline
    \end{tabular}
    \label{table:ablation}
\end{table}

\begin{figure}
\centering
\includegraphics[width=0.8\textwidth]{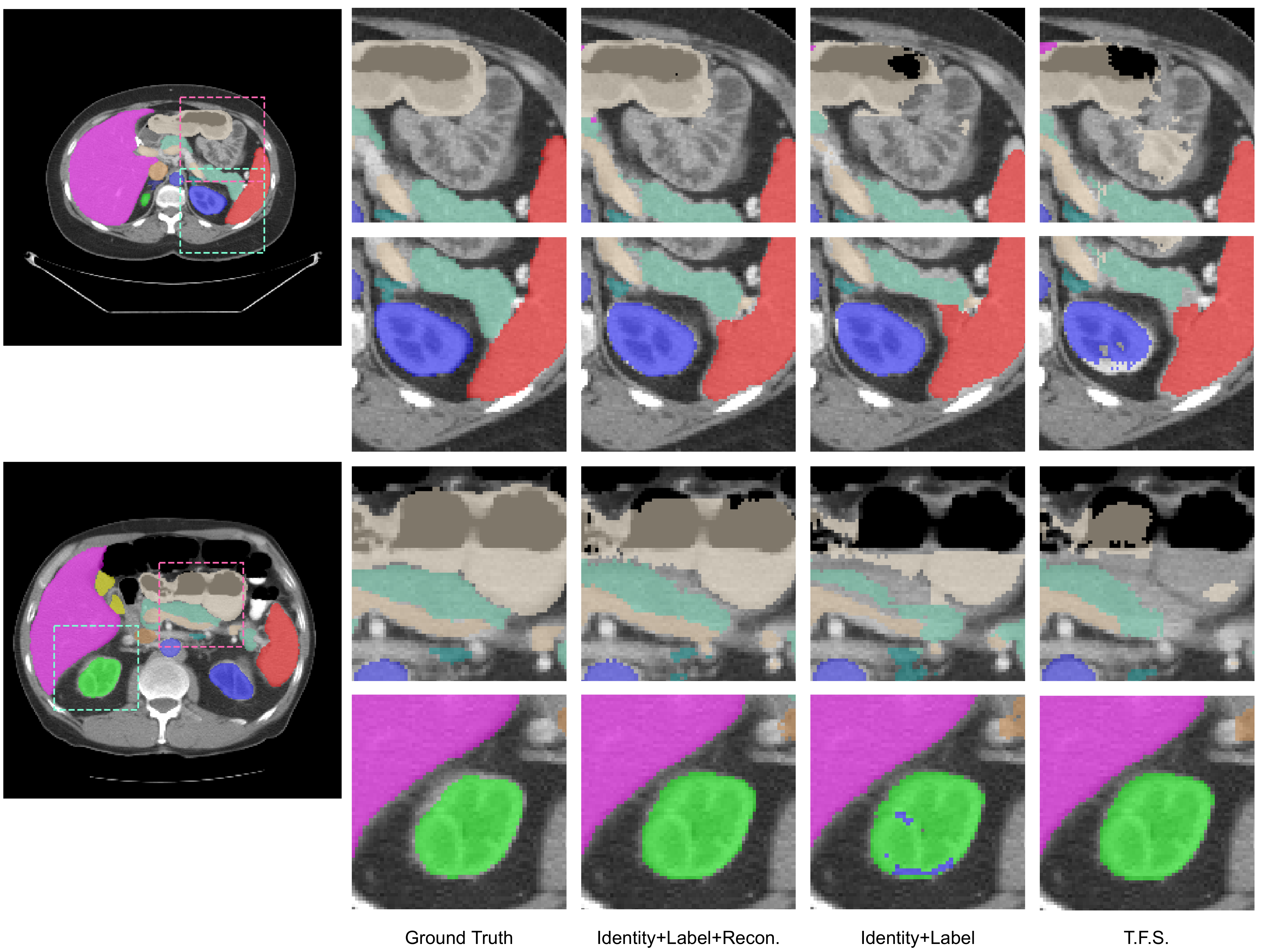}
\caption{Qualitative visualizations of ablation experiments. Regions of evident improvements are zoomed in. Better segmentation details are observed on stomach(light brown), kidney (green and blue), pancreas(cyan) and spleen (red).
} \label{fig:visualize}
\end{figure}

\subsubsection{Ablation study.}
To further analyze our method, an ablation study is conducted to characterize the roles of each module. Current results are shown in Table~\ref{table:ablation}, including experiments on 30\% labeled data of Spleen and 100\% labeled data on BTCV. The pre-trained model brings notable improvements via contrastive learning, including identity and label consistency constraints. Additionally, qualitative visualizations of experiments on BTCV are presented in Fig~\ref{fig:visualize}. Better details on stomach, kidney, pancreas and spleen are observed, demonstrating the abdominal organ segmentation improvement. The pre-trained weights learn useful representations to downstream segmentation tasks, and the incorporation of reconstruction further boosts the metric on target tasks. Besides, the performance lift on fine-tuning task demonstrates that the aggregation of labels from multiple datasets does not confuse the learning of a final model. It proves that MixCL is capable of handling the conflicts of diverse annotations and meets the intended purpose.

\section{Conclusion and Future work}
Again, data is the footstone of deep learning. The unlabeled data matters, and the data with different annotations also matters. The efforts on guiding models with knowledge in various areas, not just single, might be the start of strong artificial intelligence. We make an attempt on it, and propose Mixed contrastive learning pre-training together with various labels. Experiments introduced in this paper can preliminarily validate the performance on medical segmentation datasets. 
In future, we will continue exploring a better solution to the Problem~\eqref{problem defination} by mining more information from these pooled datasets. 
\newpage
\bibliographystyle{splncs04}
\bibliography{reference}

\begin{thebibliography}{10}
\providecommand{\url}[1]{\texttt{#1}}
\providecommand{\urlprefix}{URL }
\providecommand{\doi}[1]{https://doi.org/#1}

\bibitem{antonelli2021medical}
Antonelli, M., Reinke, A., Bakas, S., Farahani, K., Landman, B.A., Litjens, G.,
  Menze, B., Ronneberger, O., Summers, R.M., van Ginneken, B., et~al.: The
  medical segmentation decathlon. arXiv preprint arXiv:2106.05735  (2021)

\bibitem{bhalodia2020self}
Bhalodia, R., Kavan, L., Whitaker, R.T.: Self-supervised discovery of
  anatomical shape landmarks. In: International Conference on Medical Image
  Computing and Computer-Assisted Intervention. pp. 627--638. Springer (2020)

\bibitem{chen2020simple}
Chen, T., Kornblith, S., Norouzi, M., Hinton, G.: A simple framework for
  contrastive learning of visual representations. In: International conference
  on machine learning. pp. 1597--1607. PMLR (2020)

\bibitem{chen2020big}
Chen, T., Kornblith, S., Swersky, K., Norouzi, M., Hinton, G.E.: Big
  self-supervised models are strong semi-supervised learners. Advances in
  neural information processing systems  \textbf{33},  22243--22255 (2020)

\bibitem{dosovitskiy2020image}
Dosovitskiy, A., Beyer, L., Kolesnikov, A., Weissenborn, D., Zhai, X.,
  Unterthiner, T., Dehghani, M., Minderer, M., Heigold, G., Gelly, S., et~al.:
  An image is worth 16x16 words: Transformers for image recognition at scale.
  arXiv preprint arXiv:2010.11929  (2020)

\bibitem{gidaris2018unsupervised}
Gidaris, S., Singh, P., Komodakis, N.: Unsupervised representation learning by
  predicting image rotations. arXiv preprint arXiv:1803.07728  (2018)

\bibitem{grill2020bootstrap}
Grill, J.B., Strub, F., Altch{\'e}, F., Tallec, C., Richemond, P., Buchatskaya,
  E., Doersch, C., Avila~Pires, B., Guo, Z., Gheshlaghi~Azar, M., et~al.:
  Bootstrap your own latent-a new approach to self-supervised learning.
  Advances in Neural Information Processing Systems  \textbf{33},  21271--21284
  (2020)

\bibitem{hatamizadeh2022unetr}
Hatamizadeh, A., Tang, Y., Nath, V., Yang, D., Myronenko, A., Landman, B.,
  Roth, H.R., Xu, D.: Unetr: Transformers for 3d medical image segmentation.
  In: Proceedings of the IEEE/CVF Winter Conference on Applications of Computer
  Vision. pp. 574--584 (2022)

\bibitem{he2020momentum}
He, K., Fan, H., Wu, Y., Xie, S., Girshick, R.: Momentum contrast for
  unsupervised visual representation learning. In: Proceedings of the IEEE/CVF
  conference on computer vision and pattern recognition. pp. 9729--9738 (2020)

\bibitem{heller2019state}
Heller, N., Isensee, F., Maier-Hein, K.H., Hou, X., Xie, C., Li, F., Nan, Y.,
  Mu, G., Lin, Z., Han, M., et~al.: The state of the art in kidney and kidney
  tumor segmentation in contrast-enhanced ct imaging: Results of the kits19
  challenge. arXiv preprint arXiv:1912.01054  (2019)

\bibitem{hu2021self}
Hu, C., Li, C., Wang, H., Liu, Q., Zheng, H., Wang, S.: Self-supervised
  learning for mri reconstruction with a parallel network training framework.
  In: International Conference on Medical Image Computing and Computer-Assisted
  Intervention. pp. 382--391. Springer (2021)

\bibitem{liu2020self}
Liu, F., Jonmohamadi, Y., Maicas, G., Pandey, A.K., Carneiro, G.:
  Self-supervised depth estimation to regularise semantic segmentation in knee
  arthroscopy. In: International Conference on Medical Image Computing and
  Computer-Assisted Intervention. pp. 594--603. Springer (2020)

\bibitem{loshchilov2017decoupled}
Loshchilov, I., Hutter, F.: Decoupled weight decay regularization. arXiv
  preprint arXiv:1711.05101  (2017)

\bibitem{lu2020white}
Lu, Q., Li, Y., Ye, C.: White matter tract segmentation with self-supervised
  learning. In: International Conference on Medical Image Computing and
  Computer-Assisted Intervention. pp. 270--279. Springer (2020)

\bibitem{pathak2016context}
Pathak, D., Krahenbuhl, P., Donahue, J., Darrell, T., Efros, A.A.: Context
  encoders: Feature learning by inpainting. In: Proceedings of the IEEE
  conference on computer vision and pattern recognition. pp. 2536--2544 (2016)

\bibitem{roth2016data}
Roth, H.R., Farag, A., Turkbey, E., Lu, L., Liu, J., Summers, R.M.: Data from
  pancreas-ct. the cancer imaging archive. IEEE Transactions on Image
  Processing  (2016)

\bibitem{sahasrabudhe2020self}
Sahasrabudhe, M., Christodoulidis, S., Salgado, R., Michiels, S., Loi, S.,
  Andr{\'e}, F., Paragios, N., Vakalopoulou, M.: Self-supervised nuclei
  segmentation in histopathological images using attention. In: International
  Conference on Medical Image Computing and Computer-Assisted Intervention. pp.
  393--402. Springer (2020)

\bibitem{shimoda2019self}
Shimoda, W., Yanai, K.: Self-supervised difference detection for
  weakly-supervised semantic segmentation. In: Proceedings of the IEEE/CVF
  International Conference on Computer Vision. pp. 5208--5217 (2019)

\bibitem{wang2021exploring}
Wang, W., Zhou, T., Yu, F., Dai, J., Konukoglu, E., Van~Gool, L.: Exploring
  cross-image pixel contrast for semantic segmentation. In: Proceedings of the
  IEEE/CVF International Conference on Computer Vision. pp. 7303--7313 (2021)

\bibitem{windsor2021self}
Windsor, R., Jamaludin, A., Kadir, T., Zisserman, A.: Self-supervised
  multi-modal alignment for whole body medical imaging. In: International
  Conference on Medical Image Computing and Computer-Assisted Intervention. pp.
  90--101. Springer (2021)

\bibitem{xie2021propagate}
Xie, Z., Lin, Y., Zhang, Z., Cao, Y., Lin, S., Hu, H.: Propagate yourself:
  Exploring pixel-level consistency for unsupervised visual representation
  learning. In: Proceedings of the IEEE/CVF Conference on Computer Vision and
  Pattern Recognition. pp. 16684--16693 (2021)

\bibitem{xu2021deformed2self}
Xu, J., Adalsteinsson, E.: Deformed2self: Self-supervised denoising for dynamic
  medical imaging. In: International Conference on Medical Image Computing and
  Computer-Assisted Intervention. pp. 25--35. Springer (2021)

\bibitem{yang2021self}
Yang, P., Hong, Z., Yin, X., Zhu, C., Jiang, R.: Self-supervised visual
  representation learning for histopathological images. In: International
  Conference on Medical Image Computing and Computer-Assisted Intervention. pp.
  47--57. Springer (2021)

\bibitem{yao2021one}
Yao, Q., Quan, Q., Xiao, L., Kevin~Zhou, S.: One-shot medical landmark
  detection. In: International Conference on Medical Image Computing and
  Computer-Assisted Intervention. pp. 177--188. Springer (2021)

\bibitem{zbontar2021barlow}
Zbontar, J., Jing, L., Misra, I., LeCun, Y., Deny, S.: Barlow twins:
  Self-supervised learning via redundancy reduction. In: International
  Conference on Machine Learning. pp. 12310--12320. PMLR (2021)

\bibitem{zhang2021self}
Zhang, R., Liu, S., Yu, Y., Li, G.: Self-supervised correction learning for
  semi-supervised biomedical image segmentation. In: International Conference
  on Medical Image Computing and Computer-Assisted Intervention. pp. 134--144.
  Springer (2021)

\bibitem{zheng2021hierarchical}
Zheng, H., Han, J., Wang, H., Yang, L., Zhao, Z., Wang, C., Chen, D.Z.:
  Hierarchical self-supervised learning for medical image segmentation based on
  multi-domain data aggregation. In: International Conference on Medical Image
  Computing and Computer-Assisted Intervention. pp. 622--632. Springer (2021)

\bibitem{zhou2021review}
Zhou, S.K., Greenspan, H., Davatzikos, C., Duncan, J.S., Van~Ginneken, B.,
  Madabhushi, A., Prince, J.L., Rueckert, D., Summers, R.M.: A review of deep
  learning in medical imaging: Imaging traits, technology trends, case studies
  with progress highlights, and future promises. Proceedings of the IEEE
  (2021)

\bibitem{zhou2019handbook}
Zhou, S.K., Rueckert, D., Fichtinger, G.: Handbook of medical image computing
  and computer assisted intervention. Academic Press (2019)

\bibitem{zhu2020rubik}
Zhu, J., Li, Y., Hu, Y., Ma, K., Zhou, S.K., Zheng, Y.: Rubik’s cube+: A
  self-supervised feature learning framework for 3d medical image analysis.
  Medical image analysis  \textbf{64},  101746 (2020)

\end{thebibliography}
\end{document}